\documentclass[a4paper,fleqn,usenatbib]{mnras}

\usepackage[T1]{fontenc}
\usepackage{ae,aecompl}

\usepackage{graphicx}   
\usepackage{amsmath}    
\usepackage{amssymb}    

\usepackage{subfig}
\usepackage{ulem}
\usepackage{cleveref}
\makeindex \citeindextrue

\newcommand{\n}{...}
\newcommand{\rmu}{rad\,m$^{-2}$}
\newcommand{\dmu}{pc\,cm$^{-3}$}
\newcommand{\psr}{PSR\,J1745$-$2900}
\newcommand{\sgr}{\mbox{Sgr~A$^\ast$}}

\title[Polarized study of \psr\ at 7~mm]{Rotation Measure synthesis study and polarized properties of \psr\ at 7~mm}
\author[E. V. Kravchenko et al.]{E. V. Kravchenko$^{1,2}$\thanks{Contact e-mail: \href{mailto:evgenia.v.kravchenko@gmail.com}{evgenia.v.kravchenko@gmail.com}}, 
W.D. Cotton$^{3}$, F. Yusef-Zadeh$^{4}$ and Y. Y. Kovalev$^{1}$\\
$^{1}$Lebedev Physical Institute, Astro Space Center, Profsoyuznaya 84/32, Moscow 117997, Russia\\
$^{2}$Summer student at the National Radio Astronomy Observatory\\
$^{3}$National Radio Astronomy Observatory, 520 Edgemont Rd., Charlottesville, VA, 22903, USA\\
$^{4}$CIERA, Department of Physics and Astronomy, Northwestern University, Evanston, IL 60208, USA\\}

\date{Accepted 2016 February 05. Received 2016 February 03; in original form 2015 September 03}
\pagerange{\pageref{firstpage}--\pageref{lastpage}} 
\pubyear{2016}

\begin{document}
\label{firstpage}
\maketitle

\begin{abstract}
We present results of interferometric polarization observations of the recently discovered 
magnetar J1745$-$2900 in the vicinity of the Galactic center. 
The observations were made with the Karl G. Jansky Very Large Array (VLA) on 21 February 2014 in the range 40--48~GHz.
The full polarization mode and A configuration of the array were used.
The average total and linearly polarized flux density of the pulsar amounts to $2.3\pm0.31$~mJy\,beam$^{-1}$ and $1.5\pm0.2$~mJy\,beam$^{-1}$, respectively.
Analysis shows a rotation measure (RM) of $(-67\pm3)\times10^3$~\rmu, which is in a good agreement with previous measurements at longer wavelengths.
These high frequency observations are sensitive to RM values of up to $\sim2\times10^7$~\rmu.
However, application of the Faraday RM synthesis technique did not reveal other significant RM components in the pulsar emission.
This supports an external nature of a single thin Faraday-rotating screen which should be located close to the Galactic center.
The Faraday corrected intrinsic electric vector position angle is 16$\pm$9 deg East of North, and coincides with 
the position angle of the pulsar's transverse velocity.
All measurements of the pulsar's RM value to date, including the one presented here, well agree within errors, which points towards a steady nature of the Faraday-rotating medium.
\end{abstract}


\section{Introduction}
\label{s:intro}

Recently pulsar J1745$-$2900 was discovered by the \textit{Swift} satellite as an X-ray flare \citep{kennea_etal13}, 
coming from the direction of the source Sagittarius~A*~(\sgr). 
It was confirmed as a pulsar by the NuSTAR \citep{mori_etal13} and \textit{Chandra} \citep{rea_etal13} satellites.
The X-ray flaring activity and spectral properties, pulsation behavior and spin down rate measurements~\citep{gotthelf_etal13} imply PSR J1745$-$2900's magnetar nature~\citep{mori_etal13}, with a dipole magnetic field of order $10^{14}$~G.
Subsequent to the X-ray outburst, \psr\ was detected at radio wavelengths with many ground--based radio telescopes \cite[][]{buttu_etal13,eatough_etal13b,lee_etal13,shann_john13}).
\textit{Chandra} observations have shown \citep{rea_etal13} that the pulsar is located about $\sim3$~arcseconds 
away from \sgr, this is about 0.1~pc in projection (with the distance to the \sgr\ of 8.5~kpc).
Recently \cite{bower_etal15} have shown that \psr\ may be bound to \sgr\ by measuring its proper motion relative to the Galactic center (GC). 
Considering the NE2001 density model of the Galaxy \citep{NE2001}, the location of the magnetar should be less than 10~pc from the GC.

Among all known pulsars \psr\ has the largest dispersion measure, $\mathrm{DM} = 1778 \pm 3$~\dmu, 
and rotation measure, $\mathrm{RM} = -(66960 \pm 50)$~\rmu\ \citep[][]{eatough_etal13}.
\sgr\ itself has the largest observed rotation measure in the Galaxy \citep{bower_etal03,marrone_etal06,JP_etal06}, 
$\mathrm{RM} = -(5 \pm 1)\times10^5$~\rmu, which could be produced in the warm magnetized plasma, 
accreting onto the central supermassive black hole~\citep{reid_brunthaler_04,ghez_etal08,gillessen_etal09}.

\begin{figure*}
\centering
\includegraphics[angle=-90,width=0.6\textwidth]{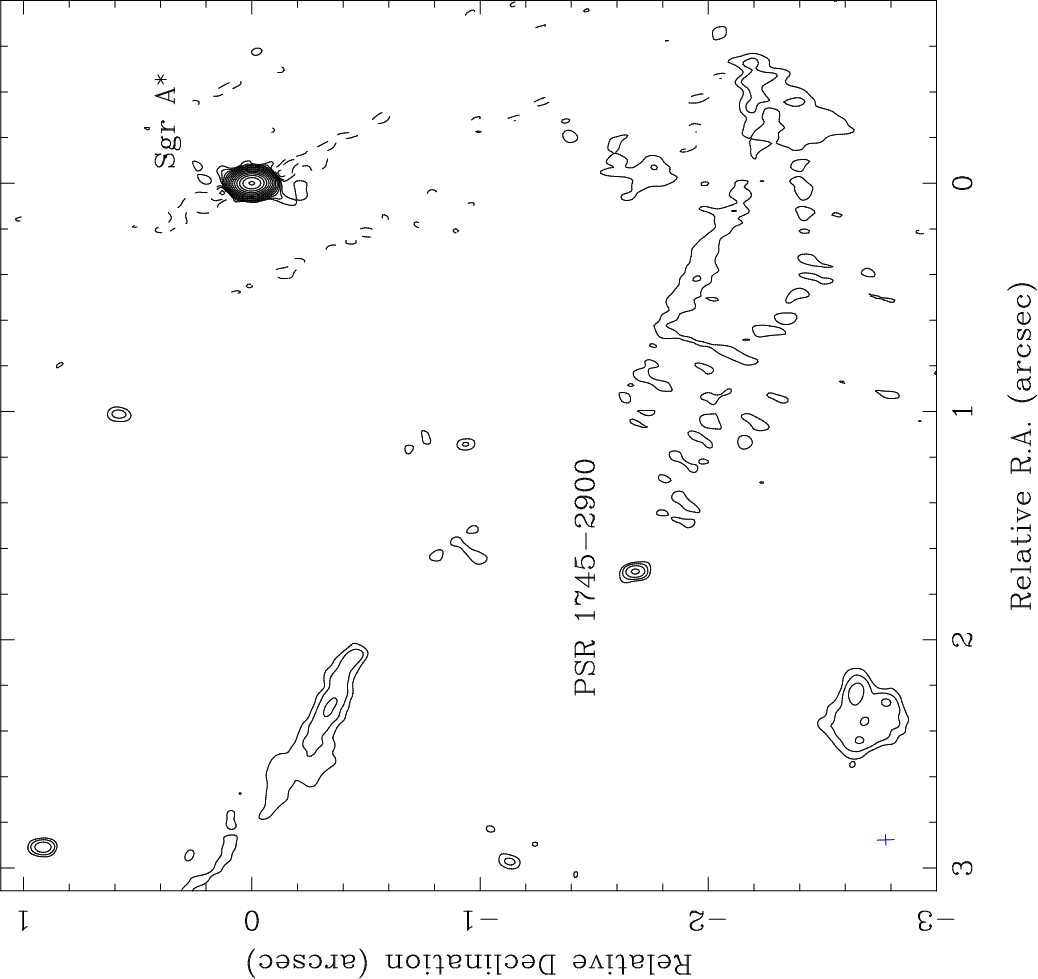}
\caption{The image of the region around \sgr with \psr, integrated over the 40--48~GHz band.
The image center is located at $\alpha\, = 17^h 45^m 40^s.038$, $\delta\, = -29^{\circ} 00' 28.069''$ (J2000).
The contours of total intensity are 0.139~mJy\,beam$^{-1}\times$($-$1, 1, 2, 4, ..., 8192).
The peak flux density of the image is of $1248.49\pm0.14$~mJy.
The half power width of the synthesized beam is 73$\times$46~milliarcseconds, the position
angle is 2$\fdg$43, and is shown at lower left corner of the image.
\label{f:image}}
\end{figure*}

So far only \psr\ and the Galactic center show such extremely large rotation measures in the Galaxy. 
Other high values are seen in non-thermal filaments $0\fdg5$ away from the GC \citep[e.g., G359.54+0.18,][]{yz_etal97} 
with maximum $\mathrm{RM}=-4200$~\rmu. 
Since RM originates in a magnetized plasma, the most probable origin of this rotation is a thin screen 
with large magnetic field, located close to the GC.
A few pulsars also show comparable RMs; two of the highest are 
PSR J1841$-$0500 with $(-2990\pm50)$~\rmu\ \citep{camilo_etal12}
and PSR J1410$-$6132 with $(2400\pm30)$~\rmu\ \citep{obrien_etal08}.
A galactic survey of neutral hydrogen at 21~cm reveals sources located in the Galactic plane with RMs of order 
$10^3$~\rmu~\citep{brown_etal07}.

Thus, the nature and location of the Faraday-rotating (also referred here to as 'Faraday screen') and scattering screens for \sgr\ and \psr\ are not ultimately known.
\cite{bower_etal14} and \cite{spitler_etal14} locate the thin scattering screen at $5.8\pm0.3$~kpc from the GC on the basis of a joint analysis of an angular broadening and temporal scattering data obtained from the pulsar.
This was also later confirmed by \cite{wucknitz14}.
At the same time, \cite{pk15} point to problems with the scenario of a scattering screen located so far away from the GC.
\cite{bower_etal03,shann_john13} explain the large \sgr\ RM as being due to a dense halo around, but not associated with the source itself. 
\cite{marrone_etal06} attribute the Faraday rotation to the material very close to \sgr\ (within $\sim0.04$~pc).
\cite{JP_etal06} show that the \sgr\ screen is external to the emission region, but also put the Faraday-rotating medium in close proximity to the GC.
\cite{eatough_etal13} place the pulsar Faraday screen within a parsec from \sgr.
The above mentioned works agree that the magnetic field in the Faraday screen should be high, with the value of tens of microgauss to a few milligauss (e.g., \cite{crocker_etal2010,noutsos_12}).
For the observed range of spectral indices, $\alpha\sim 0.2-0.3$, based on radio continuum observations of the inner few hundred pc of the Galactic center, the equipartition magnetic field is estimated to be $\sim 20\mu$G~\citep{yusef_zadeh_etal13}.

\cite{yusef_zadeh_etal15} note fluctuations in the flux density of the magnetar and suggest, that they come from interactions of the shock originated by the X-ray pulsar outburst, colliding with the orbiting GC ionized gas. Meanwhile the ionized gas acts as a Faraday-rotating screen, resulting in \psr's RM. 

\cite{eatough_etal13,shann_john13} have studied the polarization properties of the pulsar including Faraday rotation.
\cite{eatough_etal13} have measured the pulsar's RM at 2.5-8.7~GHz with the VLA, Effelsberg and Nan\c{c}ay radio telescopes to be $-(66960\pm50)$~\rmu.
At the same time, \cite{shann_john13} presented results from ATCA observations at 16-18~GHz of $-(67000\pm500)$~\rmu.
The pulsar has been observed at 42--44~GHz \citep{farhad_etal14,bower_etal15,yusef_zadeh_etal15} and at 2.5--225~GHz~\citep{torne_etal15}, but the authors did not discuss polarization. 
Here we report first results of a polarization study in the 40-48~GHz band of \psr\ at the VLA.

\section{Observations and data processing}

Observations were made with the Very Large Array of the National Radio Astronomy Observatory on 21~February~2014.
The A~configuration of the array was used, with the longest baseline reaching 36.4~km.
The field of view of the resulting image covered about 15$\times$15 arcsec centered on the Galactic center. 
Observations were done during 5 hours at frequencies from 40 to 48~GHz in full polarization mode.
Since observations were primarily targeted towards \sgr, data reduction was done following the calibration procedure 
for sources with a continuum spectrum. 
Data processing and imaging were performed within the Obit package\footnote{\url{http://www.cv.nrao.edu/~bcotton/Obit.html}} \citep{cotton08}. 
The 8192~MHz band was divided in to 4096~frequency channels with the spectral resolution of 2~MHz. 
The calibrated data were averaged over the whole 5~hr interval, 4096~frequency channels were imaged independently, 
with the phase and amplitude self-calibration application to the data.
The resulting resolution was $73\times46$ milliarcseconds, with the position angle of the synthesized beam of 2$\fdg$4.
The amplitude calibration accuracy approaches 5~per\,cent.

Instrumental polarization calibration was done using 3C286, J1733$-$1304 and J1744$-$3116 as calibrators, meanwhile absolute 
electric vector position angle (EVPA) calibration was done using 3C286 only. 
The accuracy of the instrumental polarization was of order one percent and that of the EVPA $2\fdg3$; the absolute EVPA calibration errors is estimated at $4\fdg1$.
The final calibration error of absolute vector position angle comprises errors from the instrumental and absolute calibrations and is estimated to be $4\fdg7$.

Due to the low flux density level of the pulsar and since the pulsar observations were ungated, we did not detect significant polarized signal in individual channels. 
To increase the signal-to-noise ratio (S/N), we used different averaging: (i) for the EVPA fitting about 104~MHz spectral resolution is used, resulted from averaging 52 spectral channels, and (ii) during the RM synthesis 34~MHz resolution is used, resulted from averaging 17 spectral channels.

\begin{figure}
\includegraphics[angle=-90,width=0.4\textwidth]{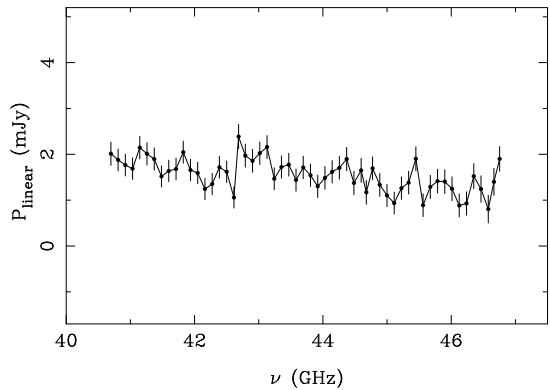}
\caption{Linear polarization at the position of \psr\ as a function of the observing frequency $\nu$. One $\sigma$ error bars are plotted, derived from noise fluctuations in the Q and U images, 
residual instrumental polarization and absolute flux calibration errors.
\label{f:pi}}
\end{figure}

\begin{figure}
\includegraphics[angle=-90,width=0.41\textwidth]{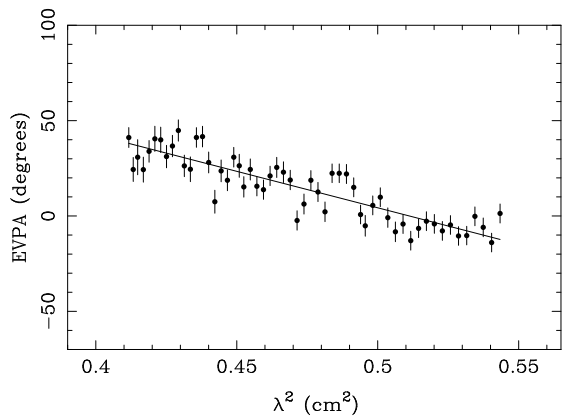}
\caption{Electric vector position angle versus wavelength squared for \psr. One $\sigma$ error bars include
errors from residual instrumental polarization, absolute EVPA calibration and noise fluctuations in the Q and U images.
\label{f:frm}}
\end{figure}

\section{Measurements of the Faraday rotation}

For the analysis of rotation measure we used two methods: linear fitting of the 
EVPA -- $\lambda^2$ dependence and RM synthesis.

\subsection{EVPA -- $\lambda^2$ analysis}

The rotation measure is defined as the slope of polarization angle,
$\psi$, versus wavelength squared, $\lambda^2$:
\begin{equation}
\mathrm{RM} = \frac{d\psi({\lambda}^2)}{d{\lambda}^2}.
\end{equation}
In the simplest Faraday-thin case, with a foreground rotating screen which is external to the emitting region, 
the polarization angle is linear with $\lambda^2$ and the RM is defined as a constant coefficient:
\begin{equation}\psi = {\psi}_0 + \mathrm{RM}\cdot{\lambda}^2,\end{equation}
where ${\psi}_0$ denotes the unaffected intrinsic EVPA of the source. 
The accuracy of the least-squares fit of the EVPA--$\lambda^2$ dependence can be evaluated if observations are made in wide bandwidth. 
In the Faraday-thick regime, when other Faraday 
effects \citep[e.g.,][]{burn66,sokoloff_etal98} take place, RM is a function of $\lambda$ and EVPA is not a linear function of $\lambda^2$.
Observations with wide enough bandwidth can distinguish Faraday-thin from Faraday-thick cases.

\subsection{Rotation Measure synthesis}

The RM synthesis technique \citep{burn66,BB05} is based on Fourier transformation of polarization signal
in $\lambda^2$ space and reconstruction of the Faraday RM spectrum $F(\phi)$ or the Faraday dispersion function.
The technique separates the different periodic behaviors in $\lambda^2$ and searches for multiple Faraday components.

The Faraday dispersion function at a particular depth is given by
\begin{equation}F(\phi) = K\int\limits_{-\infty}^{+\infty} pIe^{2i\psi}e^{-2i\phi{\lambda}^2}d{\lambda}^2,\end{equation}
where $p$ is the fractional linear polarization, $I$ -- total flux density, and $K$ is a normalization factor. 
Reconstruction of the RM spectrum is conducted over all possible values of Faraday depths $\phi$.

Rotation Measure synthesis uses full polarized information of the source to recover properties of the Faraday-rotating medium. 
The technique can not distinguish between multiple Faraday screens located on the same line of sight, when only the integrated value is available. 
Meanwhile emitting components with different sight lines through varying Faraday screens can be distinguished.
The technique can distinguish cases when radio emission traverses different Faraday screens or a mix of Faraday-rotating medium with emitting regions on its way to an observer. 
The emission from the \psr\ may include non pulsed, diffuse emission, originating in ionized gas orbiting around \sgr, as suggested by ~\cite{yusef_zadeh_etal15}, or may come from other sources in the direction of the pulsar.
Thus, application of the RM synthesis is justified in searching for multiple Faraday components.

The resolution in Faraday depth space (or ability to separate components) is defined as the width of 
the $\lambda^2$-spacing function at half maximum \citep{BB05},
where the $\lambda^2$-spacing function $R(\phi)$ is given by
\begin{equation}R(\phi) = K\int\limits_{-\infty}^{+\infty}e^{-2i\phi{\lambda}^2}d{\lambda}^2.\end{equation}
It is also known as RM transfer function~-- an equivalent to the dirty beam in aperture synthesis.
Meanwhile, the uncertainty of the reconstructed Faraday RM component is defined as the FWHM of the sampling function,
divided by twice the S/N \citep{feain_etal09,law_etal11}.
The maximum Faraday depth to which RM synthesis is sensitive is determined by the channel width, 
$\delta{\lambda}^2$: ${\phi}_\mathrm{max} \approx \sqrt{3}/\delta{\lambda}^2$.
It allows measuring $\sim2\times10^7$~\rmu\ with 34~MHz channel width.

\section{Results}
\subsection{Flux density and linear polarization}
The 40--48~GHz continuum image of the region near \sgr\ with \psr\ included is given in Fig.~\ref{f:image}.
Since \sgr\ is hundreds of times brighter than the pulsar, \sgr\ fluctuations cause artifacts on the map at the position of \psr. 
Thus, only integrated total flux density at 40~GHz was estimated and estimated as $2.3\pm0.3$~mJy.
Meanwhile, linearly polarized emission from the source has comparable values.
The variation of linearly polarized flux density with frequency is shown in Fig.~\ref{f:pi} and is observed 
at an average level of about $1.5\pm0.2$~mJy. 

\begin{figure}
\includegraphics[angle=-90,width=0.45\textwidth]{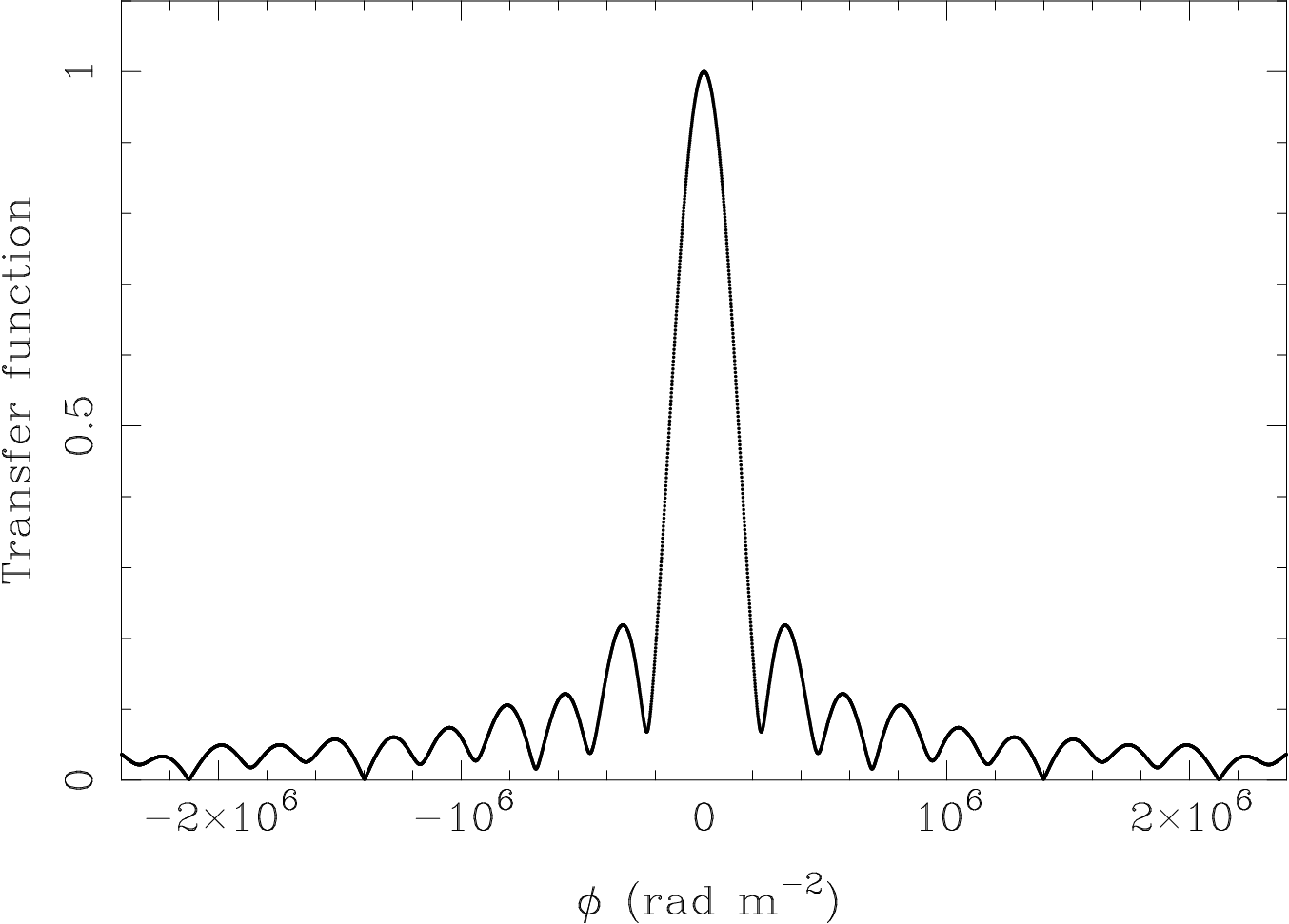}\\
\medskip
\includegraphics[angle=-90,width=0.45\textwidth]{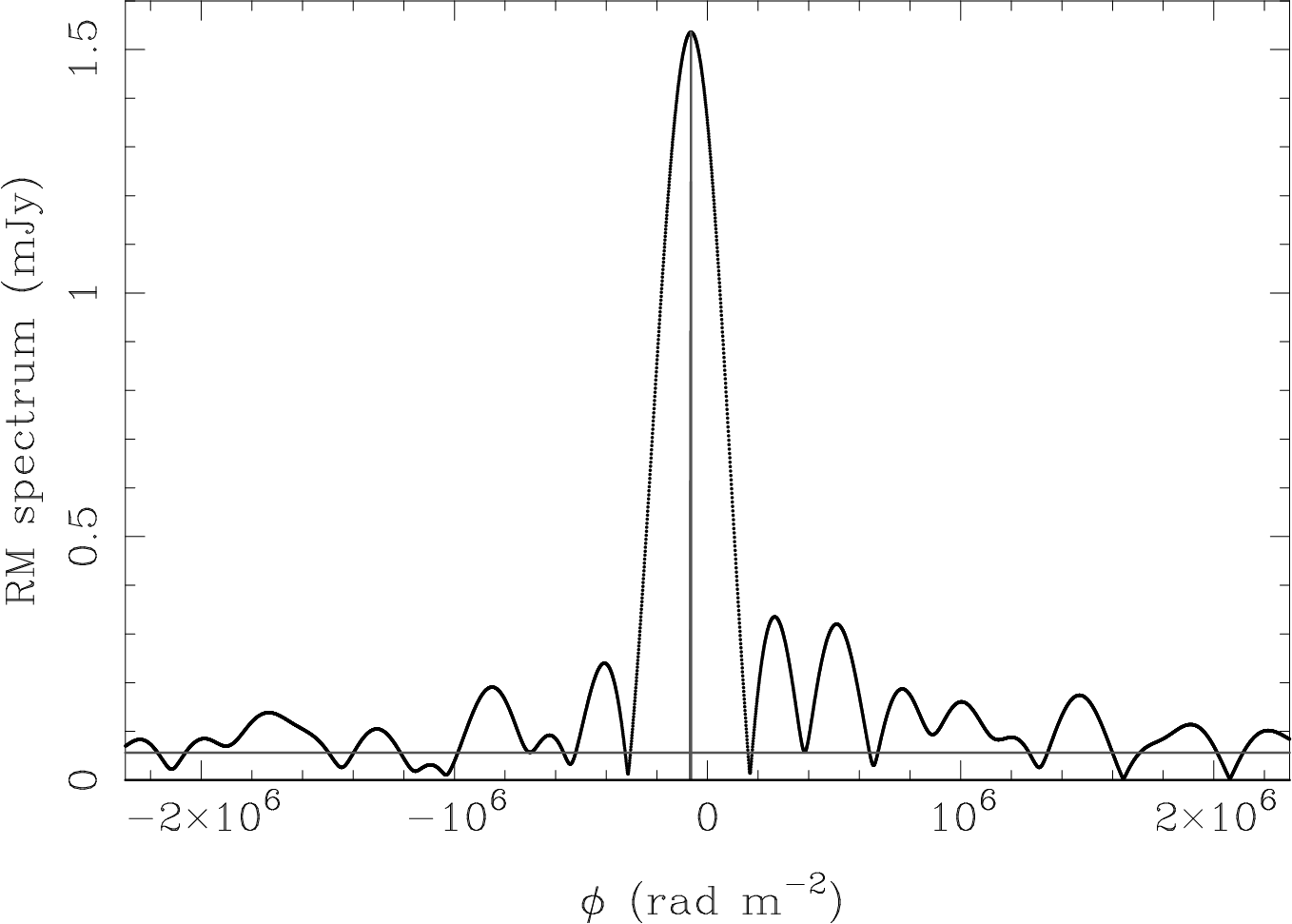}
\caption{RM transfer function given in arbitrary units (\textit{top}).
Faraday RM spectrum (\textit{bottom}) for \psr\
for a spectral resolution of 34~MHz. Vertical solid lines indicate identified Faraday components, while
the horizontal solid line represents the 1$\sigma$ rms noise level.
\label{f:rmsyn}}
\end{figure}

\subsection{Faraday rotation measure}
The behavior of EVPA with wavelength squared is shown in Fig.~\ref{f:frm} with the solid line indicating 
the weighted linear fit. The details of the fit are given in Table~\ref{t:rm}. 

In comparison to EVPA linear fitting, the higher spectral resolution of 34~MHz for the RM synthesis was used in order to increase the value of maximum detectable rotation measure.
The RM synthesis method (Fig.~\ref{f:rmsyn}) supports the results of linear fit (Table~\ref{t:rm}), though with
less accuracy. This is the result of the relatively small fractional bandwidth used in our analysis.
The amplitude of the peak in the Faraday dispersion function is $1.54\pm0.06$~mJy.
The FWHM of the sampling function is $2.45\times10^5$~\rmu, thus our observations are not sensitive to 
components with RMs differing from the main peak by less than this value.

\cite{BB05,feain_etal09,law_etal11,JP_etal12} showed that the RM synthesis technique 
is applicable to data with $\mathrm{S/N}\geq7$.
The signal is derived as the amplitude of an identified component in the Faraday RM spectrum.
The noise, $\sigma$, is taken to be average rms noise per channel divided by the square root of the number of channels used. 
The S/N for the detected RM component is 27.3. 

The measured intrinsic polarization position angle of the pulsar, $\psi_0$, is of $16\pm9$~degrees.

No other components, except the main one, were detected with the amplitude higher than $3\sigma$ in range of RMs $\pm2\times10^7$~\rmu.

\begin{table}
  \centering
  \caption{Rotation Measure results. The reduced $\chi^2$ of the fit, amplitude, $S$, its rms noise, $\sigma$, and S/N are given for the strongest component in Faraday RM spectrum. \label{t:rm}}
  \begin{tabular}{lcc}
  \hline
  &EVPA--$\lambda^2$ fit&RM Synthesis\\
  \hline
  RM (\rmu)      & $-67000\pm3000$ & $-67000\pm12000$ \\
  $\chi^2_{red}$ & 2.28            & \n               \\
  $S$ (mJy)      & \n              & 1.535            \\
  $\sigma$ (mJy) & \n              & 0.056           \\
  S/N            & \n              & 27.3             \\
  \hline
  \end{tabular}
\end{table}

We put constrains on the value of gradient in the RM, $\triangle\mathrm{RM}$, across the Faraday screen \citep{gardner_whiteoak66,tribble91}.
Assuming an  EVPA deviation of about $10^\circ$ from the linear slope as a source of beam depolarization, 
the maximum value of the gradient in Faraday screen can be estimated using the following equation
\begin{equation}
\triangle\mathrm{RM}={{\triangle \psi}\over{{\lambda}^{2}\cdot \mathrm{beam}}},
\end{equation}
with the resultant $\triangle\mathrm{RM}$ less than about 3800~\rmu.

\section{Discussion}

We measured \psr's rotation measure of $-(67000\pm3000)$~\rmu\ using the least-squares fit and $-(67000\pm12000)$~\rmu\ using the RM synthesis at 40--48~GHz. 
The resulting accuracy comes from continuum, rather than pulsed measurements.
Meanwhile our estimates are in a good agreement with other measurements at wavelengths of about and below 2~cm: $-(67000\pm500)$~\rmu\ made by \cite{shann_john13} on 1 May 2013 at 16--18~GHz
and $-(66960\pm50)$~\rmu\ made by \cite{eatough_etal13} on 28 April 2013 at 2.5--8.7~GHz.
The time separation among these and ours observations is a year, and the relatively stable RMs might indicate a steady nature of the Faraday-rotating medium. 
Though variability on time-scales of months-to-years or within estimated errors ($\pm$3269~\rmu) is possible.
The result concerning the steady nature of the Faraday-rotating medium is supported by the observations of \cite{marrone_etal06}, who showed Stokes Q and U variations of \sgr\ on the hour scale,
meanwhile the value of RM remains constant with time, and all these changes happen due to intrinsic \sgr\ activity.

The high amplitude of the RM component, almost reaching the level of the linearly polarized flux density, taken together 
with the absence of other RM components indicates that there is only one external Faraday screen, which is expected for pulsed emission.
We put an upper limit on the amplitude of a second possible component in the \psr\ RM spectrum of 0.31~mJy with RMs higher than $2.5\times10^5$~\rmu, which may originate in the unpulsed emission, probed by our continuum observations. The other imaging studies of pulsars have continuum emission, which is consistent with the pulsed, point-like emission~\citep[e.g.,][]{kaplan_etal98,kouwenhoven_00}.
The data of \cite{eatough_etal13} at 8~GHz are sensitive to RMs up to $10^7$~\rmu; they are also well described by a single RM component.

\cite{han_etal06} have reported results of polarization properties of 223 pulsars. 
Among them the PSR~J1324$-$6146 shows the highest value of $\mathrm{RM}=-1546$~\rmu\ with the $\mathrm{DM}=828$~\dmu\ 
while PSR~J1705$-$4108 has the highest $\mathrm{DM}=1077$~\dmu\ and $\mathrm{RM}=916$~\rmu.
Dispersion measures of PSR~J1705$-$4108 and \psr\ differ by a factor of 1.5 only, whereas rotation measure differ by a factor of $\sim$81.
While the dispersion measure gives the column electron density on the line of sight from the observer to the source, 
the rotation measure is also a function of the  component of the magnetic field parallel to the line of sight.
This indicates the existence of strong magnetic fields in the surrounding medium of \psr~\citep[see also discussions by][]{eatough_etal13}.
High RM values of the same sign for \psr\ and \sgr\ suggest close location of the pulsar to the GC.

The intrinsic electric vector position angle of the pulsar, $\psi_0$ is estimated to be $16\pm8$ deg. 
It coincides with the position angle of the pulsar's transverse velocity, 
which is measured to be $22\pm2$ deg \citep{bower_etal15}.
The orientation of the inferred magnetic field is consistent with a picture in which the magnetar is moving through the dense ionized medium of Sgr A West which orbits \sgr. 
The magnetar's radio emission is in part produced by synchrotron emission from electrons
accelerated in a reverse shock arising from this interaction \citep{yusef_zadeh_etal15}.
Considering the emission measure of $2\times10^7$~cm$^{-6}$pc (\cite{yusef_zadeh_etal15} and references therein) and dispersion measure of 1778~cm$^{-3}$pc, we obtain $n_\mathrm{e}\thicksim1.1\times10^4$~cm$^{-3}$ and the size of the region with the ionized medium of 0.16~pc. 
It implies the strength of magnetic field $\geq50 \mu$G.
Thus, unlike the hot X-ray emitting gas with strong magnetic field ($\geq2$~mG) within several parsecs of \sgr, proposed by \cite{eatough_etal13}, we attribute the observed pulsar's RM to the warm ionized medium with lower magnetic field, located within a few tenths of parsecs from \sgr.

So far, RMs of $10^5 - 10^7$~\rmu\ are detected only in the compact regions of active galactic nuclei
\citep[e.g.,][]{agudo_etal14,plambeck_etal14,martividal_etal15}. Such high RMs may originate in close proximity to the massive black holes powering magnetized relativistic outflows, which favors the idea that the Faraday-rotating screen of \psr\ is in the region near \sgr. Meanwhile, calculations of \cite{li_etal15} show that the main contribution to the Galactic center's RM is from accretion flow rather than a jet, if it exists. The future observations will shed light on this problem.

\section{Summary}

We observed recently discovered pulsar J1745$-$2900 with the VLA in A configuration simultaneously over the frequency range 40--48~GHz.
The traditional $\lambda^2$ fitting as well as Faraday RM synthesis were used to determine the pulsar's rotation measure.
The estimated value of ($-67\pm3$)$\times10^3$~\rmu\ agrees with results other authors previously derived at frequencies below 18~GHz.
We detected only one Faraday component in the range of $\pm2\times10^7$~\rmu\ with amplitude, approaching 100\,\% of the linearly polarized flux density.
We measured total and linearly polarized flux densities up to $2.3\pm0.3$~mJy\,beam$^{-1}$ and $1.5\pm0.2$~mJy\,beam$^{-1}$ 
levels with the VLA synthesized beam size of $73\times46$~mas, and the beam position angle of 2$\fdg$4.
These high frequency results support the external nature of an optically thin Faraday screen. 
Taken together with previous measurements, they indicate a steady nature of the Faraday-rotating medium. 
We support the view that the Faraday screen is located close to the Galactic center while \psr\ is in the vicinity of \sgr\ and attribute the observed puslar's RM to the warm ionized medium with the magnetic field of 50$\mu$G, located within a few tenths of parsecs from the GC.

\section*{Acknowledgments}
Authors would like to thank anonymous referee for useful comments and corrections, which helped us to improve the paper.
The National Radio Astronomy Observatory is a facility of the National Science Foundation operated under cooperative agreement by Associated Universities, Inc.
EVK was supported in part by the Russian Foundation for basic Research (project 14-02-31789 mol\_a).
\\
Facilities: VLA.

\bibliographystyle{mnras}
\bibliography{psr1745_mnras}

\bsp    
\label{lastpage}
\end{document}